\documentclass[pra,twocolumn,groupedaddress,showpacs]{revtex4-1}
\usepackage{graphicx,graphics,epsfig,subfigure,times,bm,bbm,amssymb,amsmath,amsfonts,mathrsfs}
\usepackage[matrix,frame,arrow]{xypic}
\usepackage[pdfstartview=FitH]{hyperref}
\usepackage{hypernat}
\usepackage{graphicx}
\usepackage{epstopdf}
\usepackage[normalem]{ulem}

\begin{document}
\title{Entanglement and its evolution following a quench in the presence of an energy current}
\author{Anirban Das$^{(1,2)}$}
\email{anirband@usc.edu}
\author{Silvano Garnerone$^{(2,3)}$}
\author{Stephan Haas$^{(1,2)}$}
\affiliation
{Departments of $^{(1)}$Physics \& Astronomy, and $^{(2)}$Center for Quantum Information Science \&
Technology, University of Southern California, Los Angeles, CA 90089, USA\\
$^{(3)}$Institute for Quantum Computing,
University of Waterloo, Waterloo, ON N2L 3G1, Canada}
\begin{abstract}
We study the Ising spin chain with a Dzyaloshinskii-Moriya interaction focusing
on the static and dynamic properties of the entanglement entropy, following a quantum quench.
We show that the effects of the additional anisotropic interaction on the phase diagram and on the
dynamics of the system are captured by the properties of the entanglement entropy. In particular,
the model provides a way to study the quench dynamics in a system with an energy current.
We consider quenches starting from an initial excited state
of the Ising spin chain, and we analyze the effects of different initial conditions.
\end{abstract}
\maketitle

\section{Introduction}
\label{sec:intro}
Recent experiments have shown that it is possible
to study unitary nonequilibrium dynamics
in quantum systems on long time scales \cite{experiments}.
One important result of these investigations has been the
observed lack of thermalization.
The reason for this non-thermal behaviour is
attributed to the near integrability of the
system. This observation
has motivated recent studies of
the non-equilibrium properties of
integrable quantum model Hamiltonians \cite{theo}.
The interest in this topic is also motivated by its
relevance to a variety of experimental situations,
including cold atoms \cite{BlDaZw}, Penning traps \cite{CiMaTo}
and Josephson-junction arrays \cite{MaScSh}.

There are many ways in which it is possible
to drive a system out of equilibrium. Quantum quenches,
and coupling to baths with different temperatures (or potentials)
are the most common protocols. While in the quench
scenario the focus is on the unitary dynamics of the full system,
in the second case one usually deals with an effective
description of the dynamics of the subsystem only.
Different out-of-equilibrium dynamics have nevertheless
similar characteristics, for example the presence of
currents (of particles, energy, or heat).

In this work, we study an Ising-like model system
driven out of equilibrium with a quantum quench in the
presence of an energy current. In the usual
quench protocol the system is prepared in the ground-state
in the absence of any current,
and subsequently the dynamics drives it to some excited
state. The model Hamiltonian we consider here allows for the
study of two different new situations: a quench from the ground-state
in the presence of an energy current, and a quench from
an initial excited state in the absence of an energy current.
The two scenarios are associated with different Hamiltonians
which, in a sense, are dual to each other.
To characterize the system's behaviour
we focus our attention on the Entanglement Entropy (EE), as measured by the Von Neumann entropy,
which is a central quantity in the characterization of nonequilibrium
quantum dynamics \cite{CaCa}.

The structure of the paper is the following: in Sec.II we introduce and describe the
properties of the model Hamiltonian, in Sec.III we consider the static properties of EE for this model,
and in Sec.IV we study the dynamics of
entanglement following a quench. Finally, we states our conclusions.
In the Appendix the reader can find more details of the calculations.

\section{The model}
We consider an Ising spin chain in transverse field $ H_I $,
with an additional Dzyaloshinskii-Moriya (DM)
interaction $ H_{DM}$. The total Hamiltonian $ H_I + H_{DM} $ is defined as follows
\begin{equation}
H=-\sum_{j}\left[\frac{1}{2}\sigma_{j}^{x}\sigma_{j+1}^{x}+\frac{h}{2}\sigma_{j}^{z}+\frac{\zeta}{8}\left(\sigma_{j}^{x}\sigma_{j+1}^{y}-\sigma_{j}^{y}\sigma_{j+1}^{x}\right)\right],
\label{Eq:ham}
\end{equation}
where $ h $ is the external magnetic field, and $ \zeta $ is the coupling parameter determining the strength
of the DM interaction. Such an anisotropic interaction is present in many low-dimensional
materials with the necessary crystal symmetry,
and it originates from spin-orbit coupling \cite{materials,dm}.
Furthermore, the DM interaction is of relevance in quantum information theory,
since it plays an important role in the
physics of quantum dots \cite{ChFrJo},
and in fault-tolerant quantum computation \cite{WuLi}.

\begin{figure}
\centering
\includegraphics[scale=0.4]{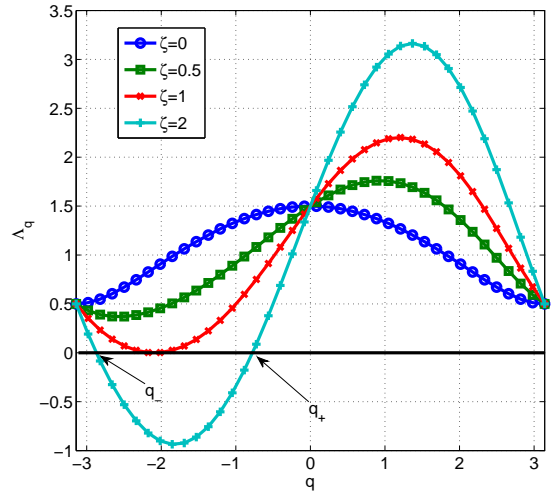}
\caption{(Color online) Spectrum of the Hamiltonian in Eq.\ref{Eq:ham}. We show the spectrum for 4 different values of $\zeta$, while keeping $h=0.5$ fixed.
See also figures in \cite{AnRaSa}.}
\label{Fig:spectrum}
\end{figure}

\begin{figure}
\centering
\includegraphics[scale=0.4]{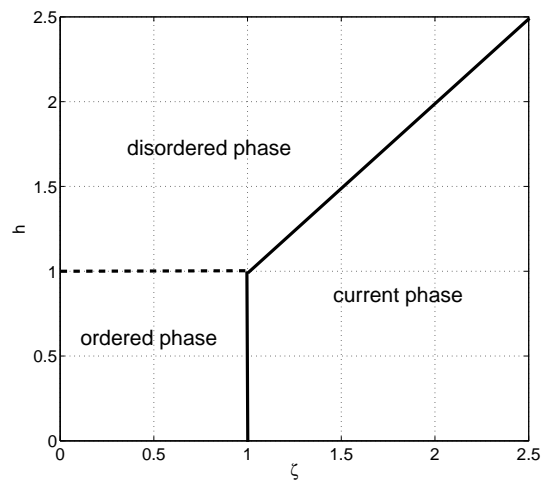}
\caption{(Color online) Phase diagram in the $h-\zeta$ plane of the model in Eq.\ref{Eq:ham}.
The dotted line is a critical line where the model shows the same universal properties of the quantum Ising model in transverse field.
See also figures in \cite{AnRaSa}.}
\label{Fig:phase}
\end{figure}

Adding the DM term to $ H_I $
does not affect the solvability of the model \cite{SiCaGa}, and interestingly
enough it provides the system with a richer phase diagram.
These features have been used in \cite{AnRaSa} to study
the effective out-of-equilibrium quantum dynamics of the model.
$ H_{DM} $ can be viewed as a current term.
The reason for this is the following.
The equation of motion for the local energy density of $ H_I $,
defined by
$
\epsilon_j = \frac{1}{2}\sigma_{j}^{x}\sigma_{j+1}^{x}+\frac{h}{2}\sigma_{j}^{z},
$
\begin{equation}
\dot{\epsilon_j} = \frac{i}{\hbar}\left[H_I,\epsilon_j \right].
\end{equation}
One can write the time derivative of the energy current as the divergence of the energy current
\begin{equation}
\dot{\epsilon_j}=C_j-C_{j+1},
\end{equation}
with
\begin{equation}
C_j \propto \sigma^{y}_{j} \left( \sigma^{x}_{j-1} - \sigma^{x}_{j+1} \right).
\end{equation}
It thus follows that $ H_{DM} $ is precisely the sum over all sites of the
local currents
 $ \sum_j C_j $.
Therefore, the ground-state expectation value of $ H_{DM} $
\begin{equation}
J \equiv
\langle \sum_j \frac{\zeta}{8}
\left(\sigma_{j}^{x}\sigma_{j+1}^{y}-\sigma_{j}^{y}\sigma_{j+1}^{x}\right)\rangle,
\end{equation}
becomes an order parameter indicating the presence of an energy current.
Once the total Hamiltonian has been diagonalized, which can be done
with the usual Jordan-Wigner and Bogoliubov transformations
\begin{equation}
H=\sum_q \Lambda_q b^\dagger_q b_q,
\end{equation}
the effect of the DM interaction is clearly observed at the single-particle level.
The single-particle spectrum is given by
\begin{equation}
\Lambda_q=\sqrt{1+h^2+2h\cos q}+\zeta \sin q,
\label{Eq:spectrum}
\end{equation}
with  $q\in [-\pi,\pi)$ the momentum of the quasi-particle.
As can be seen in the above expression and in Fig.\ref{Fig:spectrum}, the DM interaction
makes the spectrum non-symmetric with respect to $q = 0$.
Note that the ground state of the Hamiltonian including the DM interaction
is the same for all values of $ \zeta $ in the interval $ [0,1] $. In particular this
means  that, within this range of values, the Ising model in a transverse field $ H_I $
and the system described by Eq.\ref{Eq:ham} have the same ground state.
Beyond $ \zeta=1 $ (with $h\leq\zeta$) the Fermi sea starts to be populated
by the modes in between the zeros of
the single particle spectrum (i.e. between $ q_+ $ and $ q_- $ in Fig.\ref{Fig:spectrum}).
This implies that the ground state is not anymore that of $ H_I $. Furthermore,
since the DM term commutes with the rest of the Hamiltonian,
at the many-body level when $ \zeta=1 $ we must have a level crossing
between the ground state of $ H_I $ and some previously excited Hamiltonian eigenstates.
Fig.\ref{Fig:phase} shows the phase diagram of the model \cite{AnRaSa}.
There are three regions:
ferromagnetic, polarized paramagnetic, and the so-called
current phase, characterized by $ J \neq 0 $.
The current phase is gapless, and the two-point correlation functions
show a power-law behaviour with an oscillatory amplitude:
$
\langle
\sigma^x_l\sigma^x_{l+n} \rangle_{gs}
\sim\frac{Q(h,\zeta)}{\sqrt n}\cos (kn),
$
where $Q$ is a non-universal function and
$k\equiv\arccos{\frac{1}{\zeta}}$  \cite{AnRaSa}.

In the following, we study the entanglement properties of this model,
and subsequently we analyze new quench protocols for this spin system.

\section{Static entanglement entropy and phase diagram}
\label{sec:static}
In this section we show how entanglement can be used to characterize
the different phases of the model.
To measure the EE, we consider a bipartition
of the spin chain into two subsystems A and B.
For this setup a good
measure of EE between the two partitions
is given by the von Neumann entropy
$ S_A \equiv -Tr \rho_L \ln \rho_L $, where $ \rho_L $ is the ground-state
reduced density matrix of the subsystem A with $ L $ spins.

It is known that for critical one-dimensional systems
the EE scales logarithmically in the subsystem size,
with a prefactor given by the central charge of the associated Conformal Field Theory (CFT),
\begin{equation}
\label{Eq:critScaling}
S_L=\frac{c}{3}\ln L+S_0,
\end{equation}
where $ c $ is the central charge and $ S_0 $ is a non-universal constant \cite{ViLaRi,AmFaOs}.
On the other hand, in the non-critical region of the phase diagram, the entanglement entropy
saturates to a value which depends on the correlation length $ \xi $,
\begin{equation}
\label{Eq:noncritScaling}
S_L\propto \frac{c}{3}\ln \xi.
\end{equation}
Both Eq.\ref{Eq:critScaling} and Eq.\ref{Eq:noncritScaling} characterize
the ground-state properties of EE for one-dimensional systems.

Apart from the ground state it is also of interest to investigate
entanglement properties of excited states. Recently, two works have appeared
on this topic. In \cite{AlFaCa} it has been shown that
there are excited states for which
the logarithmic scaling of EE can have prefactors different from
the ground state, and that for some excited states the
scaling can be extensive in the subsystem size, instead of logarithmic.
In \cite{AlBeSi} the authors have studied
the connection
between EE for excited states and
properties of the associated CFT not contained in the central charge.
The EE of excited states is of interest also in the context of quantum quenches
since in this setting the system is unitarily driven from the initial ground state
to an excited state.

The Hamiltonian in Eq.\ref{Eq:ham} naturally fits in this set of problems.
Following the discussion from the previous section
the model we are considering allows us to study the EE
of some excited states of $ H_I $, simply
by tuning the coupling constant associated with the DM term.
The excitations we can consider in this way
are characterized by the modes in between the zeros of
the single particle spectrum that get populated when $ \zeta > 1 $ and $h < \zeta$ (\ref{Eq:spectrum}).
For these states the EE can be evaluated
analogously to the ground state of $ H_I $
(see also \cite{KaZi} for related analytical study).

\begin{figure}
\centering
\includegraphics[scale=0.45]{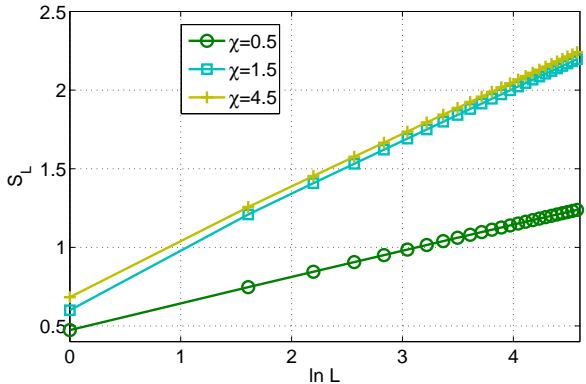}
\caption{(Color online) $S_L$ vs $L$ at $h=1.0$ for different $\zeta$. 
The scaling behavior changes from $S_L\sim\frac{1}{6}\ln L$ 
on the critical line separating the ordered and disordered phase to 
$S_L\sim\frac{1}{3}\ln L$ inside the current-carrying phase.}
\label{Fig:scaling}
\end{figure}

Let us consider in detail the entanglement properties of the different phases
shown in Fig.\ref{Fig:phase}.
First, we compare the scaling of the
entanglement in the non-current-carrying critical regions and in
the region where an energy current is present.
Fig.\ref{Fig:scaling} shows the result of the simulations for the scaling
behaviour of the ground-state EE with $\zeta<1$ and $h=1$.
For all values of $\zeta \in [0,1]$
one observes the same scaling result. In the
non-current-carrying region
critical states are present only on
the $h=1$ line of the phase diagram.
On this line, separating the ferromagnetic and polarized
paramagnetic phases,
the ground state of the system is the same as in $ H_I $.
This implies that the EE scaling is logarithmic with
a prefactor of $c/3$, and $c=1/2$.
Note that also the entire current-carrying
phase ($\zeta>1$ and $h<\zeta $) is gapless, and in this sense critical. At any point
in this phase we observe logarithmic scaling of the EE in the subsystem size.
This is consistent with the discussion in the previous section
on the algebraic decay of the two point correlation function.
Interestingly, the prefactor of the logarithmic scaling of EE in the
current phase is twice as large as the prefactor in the non-current phase.
This doubling reflects the increased number of zeros
in the single-particle spectrum, consistent with
the results of \cite{KaZi}.
In fact, when $\zeta>1$ the ground state of Eq.\ref{Eq:ham} is
the filled Fermi sea of modes in between $q_-$ and $q_+$ (see Fig.\ref{Fig:spectrum}).
Since the ground-state in the current phase is effectively
an excited eigenstate of $ H_I $, we could expect a scaling of EE
that is extensive in the system size, as shown in \cite{AlFaCa} for excited states.
The reason why this is not the case is due to the
nature of the single-particle spectrum, which at most can have two zeros (see Fig.\ref{Fig:spectrum}),
and thus does not satisfy the requirements found in \cite{AlFaCa} for an extensive scaling of EE.

\begin{figure}
\centering
\includegraphics[scale=0.40]{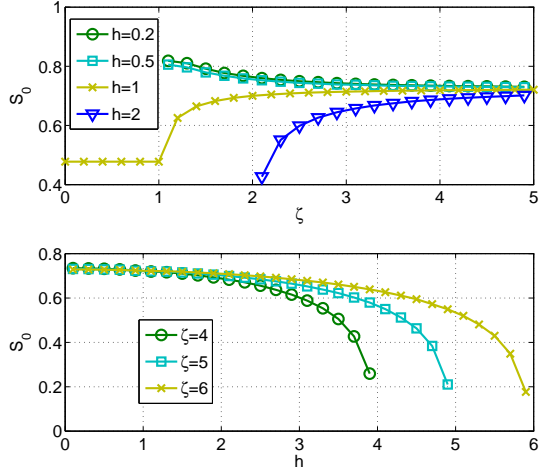}
\caption{(Color online) Non-universal nature of $S_0$. (upper panel) $S_0$ vs. $\zeta$ at different $h$;
(lower panel) $S_0$ vs. $h$ at
different $\zeta$.}
\label{Fig:sublead}
\end{figure}

The DM interaction in the Hamiltonian affects also the sub-leading term
in the scaling of EE
$ S_L=\frac{1}{3}\ln{L} +S_0(h,\zeta) $.
Deriving the analytical form of the sub-leading order term $S_0(h,\zeta)$ is
complicated. Nonetheless,
one can investigate this term numerically.
In Fig.\ref{Fig:sublead} we see that $S_0$ is
constant on the critical line $h=1$ with $\zeta\leq 1$.
As soon as $\zeta >1$ and $h<\zeta$, $S_0$ increases, but becomes almost
constant for large $\zeta$. Also $S_0$ is maximum at $h=0$,
and $S_0$ is minimum at the critical point $h=\zeta$. From the behaviour of
$S_0$ we can conclude that a given block has the highest
entanglement when all the negative modes ($q\in [-\pi,0)$) in the Fermi sea are filled.
Consequently EE increases with higher values of the energy current.

\begin{figure}
\centering
\includegraphics[scale=0.40]{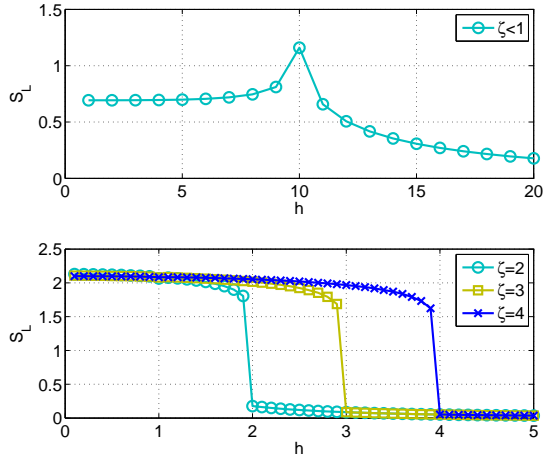}
\caption{(Color online) $S_L$ vs. $h$ at $L=60$ for different $\zeta$.
(upper panel) Static entanglement along the transition between the ordered ferromagnetic and the disordered paramagnetic phase. The peak signals
the presence of long-range correlations at the critical point, which is a signature of a second-order quantum phase transition. (lower panel) Static entanglement along the transition between the disordered paramagnetic
and the current-carrying phase. The sudden change in entanglement followed by the absence of a peak at the
critical point is a signature of the first-order quantum phase transition.}
\label{Fig:IIqpt}
\end{figure}

\begin{figure}
\centering
\includegraphics[scale=0.40]{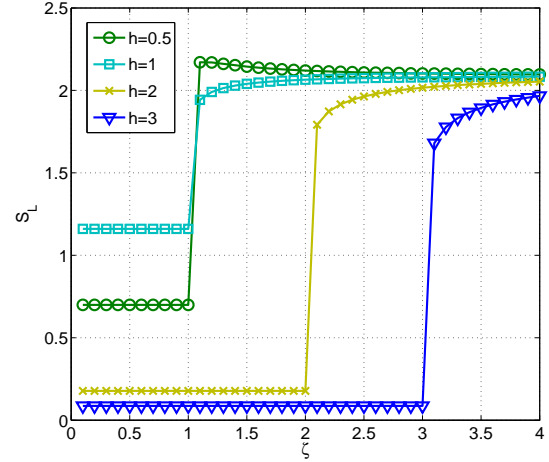}
\caption{(Color online) $S_L$ vs. $\zeta$ at $L=60$ for different $h$.
The plots for $h=0.5$ corresponds to the static entanglement
along the transition between the ordered ferromagnetic and the current-carrying phase. This is a first order
quantum phase transition which occurs at $\zeta=1$ (analogously for $h=1$).  
The plots for $h=2$ and $h=3$ correspond to the static entanglement along the transition between the disordered and
the current-carrying phase. This is a first order QPT which occurs at $\zeta=h$.}
\label{Fig:Iqpt}
\end{figure}

We now consider the differences between the critical lines shown in
the phase diagram (Fig.\ref{Fig:phase}),
separating different phases.
The only second-order quantum phase transition is found along
at the $h=1$ line (with $\zeta \leq 1$), which corresponds to the
Ising quantum phase transition
(see Fig. \ref{Fig:IIqpt}). On the hand the
boundaries of the current-carrying phase with
both the paramagnetic and the ferromagnetic phases
are characterized by a level crossing.
This translates into a sudden jump in EE (see Fig.\ref{Fig:IIqpt} and Fig.\ref{Fig:Iqpt}).
The value of EE is always higher in the current
carrying phase because of the presence of long-range correlations
that decay algebraically.
The plots in Fig.\ref{Fig:IIqpt} and Fig.\ref{Fig:Iqpt} show that controlling
the DM term can be used as an entanglement switch.
The amount of entanglement can be driven by the
DM coupling term or the magnetic field,
which are controllable parameters in
optical lattices \cite{optical}.


\section{Entanglement dynamics following a quench}
In this section, we focus on the quench dynamics
of the EE. Quenching provides a way to excite a system, initially
prepared in the ground state, and to subsequently study
the non-equilibrium dynamics of the model
(in the following, we denote
with a subscript $0$ the value of the parameters
describing the initial Hamiltonian).
As stated previously, the model in Eq.\ref{Eq:ham}
is of interest because it combines two different
mechanisms typically used to drive a system out of equilibrium:
quantum quenching, and the coupling to a field originating a current
in the system. Furthermore the inclusion of the DM term allows us to
study a model Hamiltonian where the energy current can be controlled
and used in the quench protocol.

In our setup, the quench can either involve
the magnetic field $h$, the DM coupling $\zeta$ or a combination of the two.
Since the DM term commutes with the Hamiltonian, a quench
in $ \zeta $ leaves the system in one of its eigenstates,
providing a trivial evolution of the EE.
On the other hand, quenches in the magnetic field give more interesting behaviours.
If the quench is done with the initial state prepared in a region
with no current
the results are similar to those found in \cite{CaCa}, where quenches for the
$ H_I $ were considered.
This is due to the fact that, in the absence of
a current, the ground-state wave function initially is identical
to that of $ H_I $, and the time evolution
is not affected by the presence of the DM term
(see the derivation of Eq.\ref{Eq:timeCorrelation} in the Appendix for a proof of this).
More interestingly, if the quench involves an initial state inside
the current phase, new non-trivial behaviours can be expected, since the
ground state now is radically different.
It is important to notice that the DM coupling
enters only in the specification of the initial state,
whereas the evolution can be effectively described by
the Hamiltonian without the DM term. The calculations
showing that this is in fact the case can
be found in the Appendix.

\begin{figure}
\centering
\includegraphics[scale=0.40]{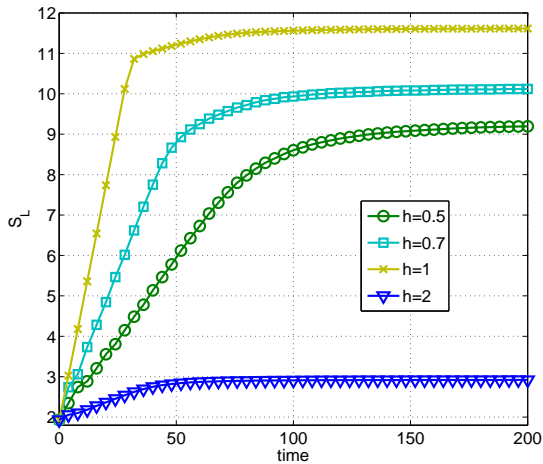}
\caption{(Color online) Quenches from the current carrying phase. 
$S_L(t)$ vs. the time steps, with $L=60$, $h_0=4.0$, $\zeta_0=\zeta=5.0$ for different $h$.
Note that the extent of the initial linear regime depends on the particular evolving Hamiltonian.}
\label{Fig:QuInCurr1}
\end{figure}
We first compare the evolution of the EE for different
quenches inside the current-carrying phase. Fixing the
coupling constant of the DM term, and quenching only the
external magnetic field we obtain the results shown in
Fig.\ref{Fig:QuInCurr1}.
One always has an initial ballistic
evolution of the EE, which grows linearly in time (measured in units where the speed of the elementary excitation is unity)
and saturates at some point. Quite interestingly,
the saturation time (hence also the rate at
which entanglement is initially building up) depends on the
particular evolving Hamiltonian.
This way we can control the time needed to
generate the maximal asymptotic amount of entanglement.
This property is relevant also from a computational point of view.
In fact, DMRG-like schemes, used for the simulation of the time evolution
of quantum systems, can take advantage of the lower rate at
which entanglement is generated. Knowing the regions in the
phase diagram where such rates are lower can provide more
efficient time simulations.
As far as we know this is a new feature that is
not present in other quench protocols considered so far
in the literature.
The other aspect that is important to notice in Fig.\ref{Fig:QuInCurr1}
is the special role played by the line $h=1$ in the phase diagram,
which turns out to provide the maximum asymptotic EE
for different quench parameters.
This can be understood by mapping the quench for
$ H_I+H_{DM} $
to a quench protocol for $ H_I $ only.
As stated in the previous section,
the entanglement evolution with respect to $H(h,\zeta)$
is identical to the evolution with respect to $H(h,0)$.
Furthermore, the ground state of $H(h_0,\zeta_0)$, the initial Hamiltonian in the quench protocol
is also an excited eigenstate of $H(h_0,0)$, because of the commutativity
of the DM term with the total Hamiltonian.
From this dual perspective
the effect of the current is that of effectively
quenching an excited eigenstate without the current term.
For the Ising model, a quench from $h_0\neq1$ yields the maximum value
of $S_L(\infty)$ when quenched to $h=1$, because the energy gap closes
at $h=1$, and hence a large number of zero energy excitations can be produced.
While the asymptotic value
of EE depends on the particular
excited state at the beginning of the quench.

\begin{figure}
\centering
\includegraphics[scale=0.40]{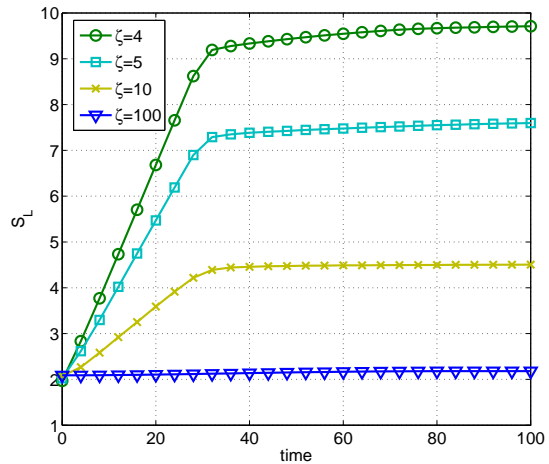}
\caption{(Color online) Quenches from the current carrying-phase with different values of the current driving field
$\zeta$. $S_L(t)$ vs. the time steps with $L=60$, $h_0=3.0$ to $h=1$ for different $\zeta_0=\zeta$.}
\label{Fig:QuInCurr2}
\end{figure}

Fig.\ref{Fig:QuInCurr2} shows results of simulations for quenches
with increasing values of the DM field in the current-carrying phase.
The asymptotic value of the EE decreases with
increasing $\zeta$. This is consistent
with the phenomenological picture provided in \cite{CaCa}, and
with the fact that if the system starts in an
excited state, the available number of unoccupied modes that
can be occupied after the quench is smaller than in the case
of having the ground state as an initial state.
Furthermore, Fig.\ref{Fig:QuInCurr2} shows that
the time at which the EE saturates does not depend on
$ \zeta $, and consequently does not depend on
the particular initial Hamiltonian eigenstates
(as long as it is not an eigenstate of the evolving Hamiltonian).
The line with $ \zeta=100 $ in Fig.\ref{Fig:QuInCurr2} shows that very deep into
the current phase quenching does not create
entanglement. In fact, when $ \zeta \gg h_0 $
quenching the magnetic field is just a small perturbation
to the Hamiltonian, which then approximately stays in
the ground state.

\begin{figure}
\centering
\includegraphics[scale=0.40]{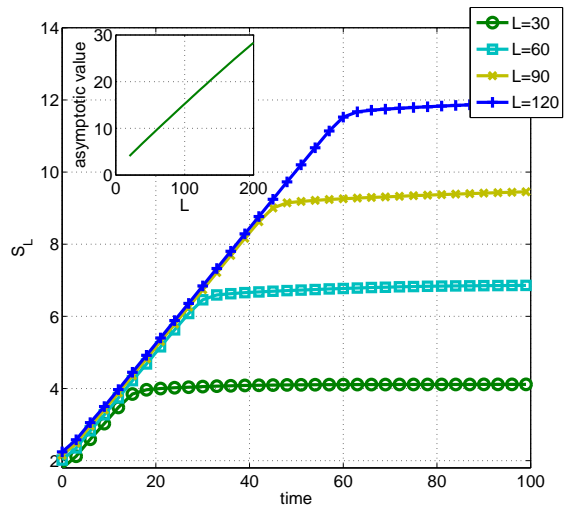}
\caption{(Color online) $S_L(t)$ vs. the time steps 
inside the current-carrying phase for different block sizes $L$. 
Quenching is done from $h_0=2.0$ to $h=1.0$ with $\zeta_0=\zeta=3.0$.}
\label{Fig:L}
\end{figure}

\begin{figure}
\includegraphics[scale=0.45]{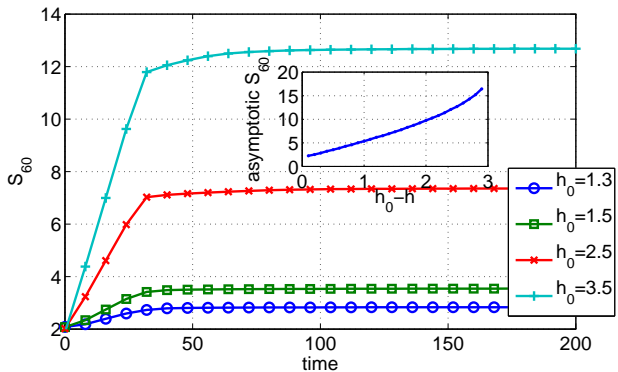}
\caption{(Color online) $S_{60}(t)$ vs. the time steps 
for quenches to $h=1$ from various $h_0$ inside the current-carrying phase at $\zeta=\zeta_0=4$.}
\label{Fig:quenchsize}
\end{figure}

Finally, we verify that the presence of
an energy current does not affect the extensive nature of the
asymptotic value of EE (Fig.\ref{Fig:L}), and
its proportionality with the quench size (Fig.\ref{Fig:quenchsize}).

\section{Conclusions}
We have studied the static and dynamic properties
of the entanglement entropy in the Ising spin chain with a transverse field and
a Dzyaloshinskii-Moriya interaction. The model is characterized
by the presence of an energy current for certain regions of the phase diagram.

Concerning the static properties we have analyzed
the transitions between phases with no energy current and the
phase where an energy current is present.
The transition is captured by a discontinuity of EE as a function of the parameters,
and by a distinguishable scaling behaviour in the current-carrying and non-current-carrying regions.
In particular, the leading logarithmic term of the
EE scaling with respect to the system size
has a prefactor in the current-carrying region which
is twice as large compared to the second order Ising critical line.

Concerning the behaviour of the entanglement evolution
following a quench, the model in Eq.\ref{Eq:ham}
allows us to study new quench protocols.
The usual schemes consider quenches from an initial ground state.
This scenario, for the model in Eq. \ref{Eq:ham},
effectively corresponds to a quench from an initially excited state
of the Ising spin chain in transverse field (without DM interaction).
The main result of this analysis shows that
the ballistic picture presented in \cite{CaCa} is still valid,
although with a significantly different aspect.
In particular the entanglement saturation time in the current-carrying
phase depends on the
details of the evolving Hamiltonian.
This is an indication of the role played by the evolving 
Hamiltonian on the propagation of
excitations. This result is of relevance in
tuning the dynamics of the system in regions with a
different rate for the propagation of entanglement.
Furthermore it also provides
a characterization of the regions in the phase diagram that can be
simulated more efficiently with DMRG-like techniques.

From a general point of view, the model in Eq. \ref{Eq:ham} also
suggests a simple way to study the quench dynamics of initial excited
states in integrable systems. The addition of a commuting term
in the Hamiltonian causes a reshuffling of the spectrum that,
without changing the integrability of the model, allows us to
obtain non-trivial results about the excitations in the original model.
The same trick can be applied to other systems of interest.

We thank Letian Ding, Zoltan R\'acz, and Paolo Zanardi for useful comments.
This work has been supported by NSF grants PHY-803304, and DMR-0804914.

\appendix
\bigskip{}

\appendix
\bigskip{}

\textbf{Appendix.}
In this Appendix we give a detailed description of the steps involved in first
evaluating the entanglement entropy of the Hamiltonian in Eq.\ref{Eq:ham},
and then calculating its time evolution.
After the standard sequence of
Jordan-Wigner and Bogoliubov transformations the Hamiltonian
is in the diagonal form $H=\sum_{k=-\pi}^\pi\Lambda_kb_k^{\dag}b_k$,
with $\Lambda_k=\frac{1}{2}(\sqrt{1+h^2+2h\cos k}+\zeta \sin k)$.\\
The density matrix of the subsystem od size L, embedded in a system of size N,
can be obtained tracing out the rest of the system
\begin{equation}
\rho_L=Tr_{N-L}(\rho)=A_0 e^{-\cal{H}},
\label{}
\end{equation}
where $A_0$ is a normalization constant and $\cal{H}$ is a quadratic hermitian operator
\begin{equation}
{\cal{H}}=\sum_{i,j=1}^{L} c^\dag_iV_{i,j}c_j+\frac{1}{2}(c^\dag_iW_{i,j}c^\dag_j-c_iW_{i,j}c_j).
\label{}
\end{equation}
$\cal H$ can be diagonalized via a generalized Bogoliubov transformation.
The reduced density matrix has the form
\begin{equation}
\rho_L=A_0\exp[-\sum_{q=1}^L\varepsilon_qd_q^\dag d_q]
\label{}
\end{equation}
Using $Tr(\rho_L)=1$, we get $A_0=\Pi_{q=1}^L\frac{1}{1+\exp(-\varepsilon_q)}$.
This gives the final form of the density matrix as
\begin{equation}
\rho_L=\Pi_{q=1}^L\frac{\exp(-\varepsilon_qd^\dag_qd_q)}{1+\exp(-\varepsilon_q)}.
\label{}
\end{equation}
Defining $\nu_q\equiv \frac{1-\exp(-\varepsilon_q)}{1+\exp(-\varepsilon_q)}$, we can write
\begin{equation}
\rho_q\equiv
\left( \begin{array}{c c}
\frac{1+\nu_q}{2} & 0\\
0 & \frac{1-\nu_q}{2}
\end{array} \right),
\label{}
\end{equation}
and also
\begin{align}
\rho_L&=\Pi_{q=1}^L\frac{1+\nu_q}{2}\exp[-\ln(\frac{1+\nu_q}{1-\nu_q})d_q^\dag d_q]\nonumber\\
&=\Pi_{q=1}^L(\frac{1+\nu_q}{2}-\nu_qd_q^\dag d_q)\nonumber\\
&=\bigotimes_{q=1}^L\rho_q.
\end{align}
Using the fact that $\exp(d_q^\dag d_q\ln \lambda)=1+(1-\lambda)d_q^\dag d_q$, one has for the entanglement entropy
\begin{align}
S_L&=-Tr(\rho_L\ln(\rho_L))\nonumber\\
&=\sum_{q=1}^L[\ln(1+\exp(-\varepsilon_q))+\frac{\varepsilon_q}{1+\exp(\varepsilon_q)}]\nonumber\\
&=-\sum_{q=1}^L(\frac{1+\nu_q}{2}\ln\frac{1+\nu_q}{2}+\frac{1-\nu_q}{2}\ln\frac{1-\nu_q}{2}).
\label{}
\end{align}
We have to calculate $\nu_q$, from which we can obtain the block entropy.
$\nu_q$ is given by the expectation value of $d_q^\dag d_q$ and $d_qd_q^\dag$
\begin{align}
\langle d_q^\dag d_q \rangle &=Tr(\rho_Ld_q^\dag d_q)=\frac{\exp(-\varepsilon_q)}{1+\exp(-\varepsilon_q)}=\frac{1-\nu_q}{2}\nonumber\\
\langle d_qd_q^\dag \rangle &=Tr(\rho_Ld_qd_q^\dag)=\frac{1}{1+\exp(-\varepsilon_q)}=\frac{1+\nu_q}{2}.
\label{}
\end{align}
We define four $2L\times 1$ column vector: $D\equiv
\left( \begin{array}{c}
d \\
d^\dag
\end{array} \right)$, $C \equiv \left( \begin{array}{c}
c \\
c^\dag
\end{array} \right)$, $\bar D\equiv \left( \begin{array}{c}
d^\dag  \\
d
\end{array}\right)$
and  $\bar C \equiv \left( \begin{array}{c}
c^\dag  \\
c
\end{array}\right),$
where $d=(d_1,\dots,d_L)^t$, and similarly for $c$.
The previous Bogoliubov transformations can be expressed in a compact matrix notation as
\begin{equation}
D=
\left( \begin{array}{c c}
g & h\\
h & g
\end{array} \right)
C,
\end{equation}
and
\begin{equation}
\bar{D}^t=\bar{C}^t
\left( \begin{array}{c c}
g^t & h^t\\
h^t & g^t
\end{array} \right),
\end{equation}
where $g$ and $h$ are L$\times$L matrices.
In terms of expectation values we have
\begin{equation}
\langle D\bar{D}^t \rangle =
\left( \begin{array}{c c}
g & h\\
h & g
\end{array} \right)
\langle C\bar{C}^t \rangle
\left( \begin{array}{c c}
g^t & h^t\\
h^t & g^t
\end{array} \right).
\label{}
\end{equation}

Let us now consider a quantum quench protocol. Initially the system
is prepared in the ground state of an Hamiltonian $H'$, and
suddenly one of the parameters is changed, and the new Hamiltonian is
denoted by $H$.
The quasi-particle operator vector $B'_k\equiv (b'_k, \; {b'}_{-k}^\dag)^t$ is associated with
$H'$, and the vector $B_k\equiv (b_k, \; {b}_{-k}^\dag)^t$ is associated with $H$. Similarly for $C_k$,
which is associated with bare vacuum fermions.
Define also the matrix $R_\mu(\alpha) \equiv \cos(\frac{\alpha}{2}) \mathbb{I}+i\sigma_\mu\sin(\frac{\alpha}{2})$, where
$\sigma_\mu$ are the Pauli matrices, and $\mu=x,y,z$.
It can be easily seen that $C_k=R_x(\theta_k)B_k$, and $C_k=R_x(\theta'_k)B'_k$, with $\theta_k$ a parameter
of the Bogoliubov transformation \cite{SePoSa} . From which we can write
$B_k=R_x(\theta'_k-\theta_k)B'_k$.

When a quench takes place, the time evolution of the fermion operators is given by $B_k(t)=e^{-iHt}B_ke^{iHt}$.
We can write $B_k(t)=S_z(-2\Lambda_kt)B_k$, where
$$S_z(-2\Lambda_kt)=
\left( \begin{array}{c c}
e^{-i\Lambda_kt} & 0\\
0 & e^{-i\Lambda_{-k}t}
\end{array} \right).$$
Notice that the energy spectrum of the Hamiltonian in Eq.\ref{Eq:ham}
is not symmetric, which means that in general $\Lambda_{-k} \neq \Lambda_k$.
In order to evaluate the two-point correlation functions we consider
different cases. When the initial state of the system is in the non-current-carrying region we have
\begin{equation}
\langle B'_k{B'_k}^\dag \rangle=
\left( \begin{array}{c c}
1 & 0\\
0 & 0
\end{array} \right).
\label{B1}
\end{equation}
If the initial state of the system is in the current-carrying phase and $k\in(k_1,k_2)$, where $k_1$ and $k_2$
are the zeros of the spectrum then
\begin{equation}
\langle B'_k{B'_k}^\dag \rangle=
\left( \begin{array}{c c}
0 & 0\\
0 & 0
\end{array} \right).
\label{B2}
\end{equation}
For $k$ lying between $-k_1$ and $-k_2$ we have
\begin{equation}
\langle B'_k{B'_k}^\dag \rangle =
\left( \begin{array}{c c}
1 & 0\\
0 & 1
\end{array} \right).
\label{B3}
\end{equation}
A compact way of expressing Eq.\ref{B1}, Eq.\ref{B2}, and Eq.\ref{B3} is given by
\begin{equation}
\langle B'_k{B'_k}^\dag \rangle =
\left( \begin{array}{c c}
\frac{1}{2}(1+\frac{|\Lambda_k|}{\Lambda_k}) & 0\\
0 & \frac{1}{2}(1-\frac{|\Lambda_{-k}|}{\Lambda_{-k}})
\end{array} \right).
\label{}
\end{equation}
Finally we can write
\begin{align}
&\langle C_k(t)C_k^\dag(t) \rangle =R_x(\theta_k) \langle B_k(t)B_k^\dag(t) \rangle {R_x^\dag(\theta_k)}\nonumber\\
&=R_x(\theta_k)S_z(-2\Lambda_kt) \langle B_kB_k^\dag \rangle S_z^\dag(-2\Lambda_kt){R_x^\dag(\theta_k)}\nonumber\\
&=R_x(\theta_k)S_z(-2\Lambda_kt)R_x(\theta'_k-\theta_k) \langle B'_k{B'_k}^\dag \rangle \nonumber\\
&\times R_x^\dag(\theta'_k-\theta_k)S_z^\dag(-2\Lambda_kt){R_x^\dag(\theta_k)}.
\label{Eq:timeCorrelation}
\end{align}
Notice that the above expression is the same if we consider $ S_z(-2\Lambda_kt) $, with $\Lambda_k $ the
single particle spectrum of Eq.\ref{Eq:ham}, or if we consider $ S_z(-2\Lambda_kt) $, with $\Lambda_k $
the single particle spectrum of the Ising Hamiltonian without the DM term. This can be seen with a direct calculation.
For example, one entry of the above correlation matrix is given by
\begin{align}
&2 \langle c_{-k}^\dag(t) c_{-k}(t) \rangle = \nonumber\\
&E_1+E_2+ \left( E_2-E_1 \right) \cos {\theta_k} \cos{\left(\theta'_k-\theta_k \right)} \nonumber\\
&+\left( E_1-E_2 \right) \cos \left[ t \left( \Lambda_k+\Lambda_{-k} \right) \right] \sin{\theta_k} \sin \left(\theta'_k-\theta_k\right),
\end{align}
where $E_1 \equiv \frac{1}{2}(1+\frac{|\Lambda_k|}{\Lambda_k})$ and $E_2 \equiv \frac{1}{2}(1-\frac{|\Lambda_{-k}|}{\Lambda_{-k}})$.
The argument of $S_z$ appears only in the argument of the trigonometric function in such a way that the DM contribution is irrelevant (see
Eq.\ref{Eq:spectrum}).
This proves that the time evolution of the correlation matrix in Eq.\ref{Eq:timeCorrelation},
with respect to the ground-state of Eq.\ref{Eq:ham}, is
the same as the time evolution of the correlation matrix with respect to the Ising Hamiltonian in transverse field,
with respect to an excited state of the Ising Hamiltonian.
\end{document}